\documentclass[%
 reprint,
showpacs,preprintnumbers,
 amsmath,amssymb,
 aps,
 longbibliography,
]{revtex4-1}

\usepackage{graphicx}% Include figure files
\usepackage{dcolumn}% Align table columns on decimal point
\usepackage{bm}% bold math
\usepackage{hyperref}% add hypertext capabilities
\begin{document}

%\preprint{}

\title{Modulation of electronic and mechanical properties of phosphorene through strain}

\author{Mohammad Elahi}
\author{Kaveh Khaliji}
\author{Seyed Mohammad Tabatabaei}
 \affiliation{School of Electrical and Computer Engineering, University of Tehran, Tehran 14395-515, Iran}
\author{Mahdi Pourfath}
\email{pourfath@ut.ac.ir}
\email{pourfath@iue.tuwien.ac.at}
 \affiliation{School of Electrical and Computer Engineering, University of Tehran, Tehran 14395-515, Iran\\
 Institute for Microelectronics, Technische Universit\"at Wien, Gu{\ss}hausstra{\ss}e 27--29/E360, A-1040 Wien, Austria}
\author{Reza Asgari}
\email{asgari@ipm.ir}
 \affiliation{School of Physics, Institute for Research in Fundamental Sciences (IPM), Tehran 19395-5531, Iran}

\date{\today}
\begin{abstract}
We report a first-principles study on the elastic, vibrational, and electronic properties of the recently synthesized phosphorene. By calculating the Gr\"uneisen parameters, we evaluate the frequency shift of the Raman/infrared active modes via symmetric biaxial strain. We also study a strain-induced semiconductor-metal transition, the gap size, and the effective mass of carriers in various strain configurations. Furthermore, we unfold the emergence of a peculiar Dirac-shaped dispersion for specific strain conditions including the zigzag-oriented tensile strain. The observed linear energy spectrum has distinct velocities corresponding to each of its linear branches and is limited to the $\Gamma-X$ direction in the first Brillouin zone.
\end{abstract}
%\keywords{phosphorene, band gap, biaxial strain, Dirac dispersion, effective mass, Gr\"unesien parameter.}
\maketitle

\section{Introduction}
The discovery of graphene in 2004 has triggered an unprecedented leap in the research on ultrathin two-dimensional (2D) crystals~\cite{Science, Beyond}. Such crystals are mostly exfoliated into individual thin layers from their layered counterparts. Famous examples include graphene, hexagonal boron nitride~\cite{Geim}, and molybdenum disulfide, the latter being the most well-known member of the family of 2D transition metal dichalcogenides~\cite{wang12}. Due to the wealth of exquisite physical phenomena that arise when charge, spin and heat transport are restricted within a 2D plane, these materials have been among the most interesting subjects in condensed matter physics~\cite{Xu}.

The intriguing prospect of the potential nano-electronic applications which may take advantage of the impact of quantum confinement and dimensionality reduction in 2D materials has enticed the scientific community to actively explore possibilities of similar materials with outstanding characteristics. In this regard, phosphorene, an atomically thin layer of the element phosphorus which has a natural band gap, has been synthesized recently through mechanically cleaving bulk black phosphorus (BBP) followed by a plasma-assisted thinning process~\cite{Plasma}. In phosphorene, the atoms are arranged in a rectangular lattice with the surface being slightly puckered (see Fig.~\ref{f:Fig1}(a)), giving rise to novel correlated electronic phenomena ranging from semiconducting to superconducting behaviors. Moreover, the monolayer is still planar enough to confine electrons so that charge flows quickly, leading to a relatively high mobility that is promised by the electronic, optical, mechanical, chemical, and thermal properties~\cite{qiao2014, ref-elec2, ACSNano, ref-reviewer-1, ref-reviewer-2}. In particular, being a semiconducting 2D material, phosphorene now renders to be an appealing candidate for nano-electronic applications as asserted in field effect transistors based on multi-layers of this material~\cite{APLEx, ACSNano, Ex4, Ex5, Ex6, Deng14, Zhang14}.

The controlled introduction of strain into semiconductors offers an important degree of flexibility in both scientific and engineering applications. To gain insight on how phosphorene can be fruitful in the realization of high performing devices, fundamental studies on the strain-induced variation of mechanical and electronic properties of this material are essential. This can be readily evidenced by referring to both the ubiquity of mechanical perturbations and the numerous previous investigations regarding the possibility of amending the electronic properties of 2D materials through strain engineering~\cite{PRB, APL, APLSi, RCS, Mos2_strain, Mamad}. In this context, it has been recently proposed that via perpendicular compression, the electronic band structure of phosphorene undergoes a semiconductor-semimetal-metal transition~\cite{Castro}. Peng {\it et ~al.} has also demonstrated a strong modulation of both band gap and effective mass of carriers in response to axial in-plane deformations~\cite{Wei1}. Moreover, a unique anisotropic conductance is reported which can be controlled and even rotated by $90$ degrees under specific uniaxial and symmetric biaxial strains~\cite{Fei}.

In this paper, we carry out first-principles simulations to investigate the elastic, vibrational, and electronic properties of phosphorene. Our numerical results show a negative Poisson's ratio in the out-of-plane direction under uniaxial deformations oriented along the zigzag direction. To obtain the Gr\"uneisen parameters, the frequency shift of the Raman/infrared active modes through symmetric biaxial strain are evaluated. We demonstrate the feasibility of in-plane deformations in inducing semiconductor-metal transition and manipulating the gap size and effective mass of carriers in various strain distributions. With the application of specific biaxial strain distributions, we further report on the formation of a peculiar Dirac-like energy spectrum. The obtained electronic dispersion is queer as it is linear along the $\Gamma-X$ direction, while being parabolic along the orthogonal path.

This paper is organized as follows. Sec. II describes the methodology employed for the electronic and phononic calculations. In Sec. III, the relevant mechanical constants of phosphorene, the variation of the Raman/infrared active modes with strain, along with the strain-induced modulation of the electronic dispersion are investigated. Finally, a short summary and concluding remarks are presented in Sec. IV.
%%%%%%%%%%%%%%%%%%%%%%%%%%%%%%%%%%%%%%%%%%%%%%%%%%%%%%%%%%%%%%%%%%%%%%%%%%%%%%%%%%%%%%%%%%%%%%%%%%%%%%%%%
%%%%%%%%%%%%%%%%%%%%%%%%%%%%%%%%%%%%%%%%%%%%%%%%%%%%%%%%%%%%%%%%%%%%%%%%%%%%%%%%%%%%%%%%%%%%%%%%%%%%%%%%%
%%%%%%%%%%%%%%%%%%%%%%%%%%%%%%%%%%%%%%%%%%%%%%%%%%%%%%%%%%%%%%%%%%%%%%%%%%%%%%%%%%%%%%%%%%%%%%%%%%%%%%%%%
\section{Theory and Method}
We carry out first-principles simulations based on the density-functional theory (DFT) as implemented in the SIESTA code~\cite{Siesta1} to perceive the relevant mechanical, vibrational and electronic properties of a single layer phosphorene. The VASP package~\cite{VASP1} is also used in some instances throughout this paper to provide increased precision for critical results. Apart from the package used, calculations begin with the determination of the optimized geometry~\cite{CG}, i.e., the configuration in which the residual Hellmann-Feynman forces acting on atoms are smaller than $0.01$ eV/\AA. In the SIESTA code, this can be achieved by employing a double-$\zeta$-double-polarized (DZDP) basis set along with the conjugate gradient method within the generalized gradient approximation (GGA) formalism and taking advantage from the norm-conserving Troullier-Martins pseudo-potentials~\cite{PBE, Siesta3}. In the VASP package, the projector augmented wave method along with the Perdew-Burke-Ernzerhof (PBE) form of the exchange correlation functional are adopted for the calculation of the exchange-correlation energy~\cite{VASP2, PBE}. A cutoff energy equal to $180$ Ry ($500$ eV) is used for calculations using SIESTA (VASP) so that assure a total energy convergence better than $0.01$ meV/unit cell in obtaining the self-consistent charge density. A vacuum separation of $15$ \AA, which is sufficient to hinder interactions between adjacent layers, is adopted. Sampling of the reciprocal space Brillouin zone is done by a Monkhorst-Pack grid of $\mathrm{12\times12\times1}$ $k$-points. The phonon dispersion curves and the Raman/infrared active modes are calculated by diagonalizing the dynamical matrix obtained by the small-displacement method (SDM) with forces calculated in a $\mathrm{4\times4}$ supercell~\cite{SDM}.

In contrast to the flatness of graphene, phosphorene is a puckered honeycomb structure with each phosphorus atom covalently bonded with three neighboring atoms within a rectangular unit cell (see Fig.~\ref{f:Fig1}(a)). The crystal structure is spanned by lattice vectors $\vec{a}_{1}=a_1\widehat{x}$ and $\vec{a}_{2}=a_2\widehat{y}$ along armchair and zigzag directions, respectively. The distinct armchair ridges in the side view of phosphorene in Fig.~\ref{f:Fig1}(a) are characterized by the lattice buckling constant, i.e., $\Delta_{z}$.

We first calculate the structural parameters of BBP and a monolayer phosphorene and compare the results for BBP with those results obtained in experiment~\cite{ExpBBP}. We use, by treating van der Waals (vdW) interactions between adjacent layers in BBP, the Grimme correction to the PBE functional in SIESTA~\cite{Grimme} and thus the lattice parameters are calculated and summarized in Table~\ref{tab1}. The DZDP basis set along with the PAW pseudo-potentials are employed to calculate the same set of structural parameters. It is noted that the DZDP method, excluding the vdW treatment, provides adequate accuracy in terms of its compliance with the reported experimental values for BBP. Moreover, the tiny discrepancies between VASP and SIESTA results can be ascribed to the different parameterizations of the functionals used and to the different basis sets employed in each package (plane waves versus numerical atomic orbitals). Therefore, the DZDP basis set without the vdW correction is adopted as the main tool for evaluating the results. In some instances, the PAW method is also invoked where critical results have been encountered. It should be mentioned that our calculated structural parameters for both BBP and phosphorene are in excellent agreement with reported before calculations~\cite{ACSNano, Wei1, Wei2, vdWBBP,qiao2014}.

\begin{table}[h]
\resizebox{0.5\textwidth}{!}{%
\begin{tabular}{c|c|c|c|c|c|c|c} \hline
Material & Method & ~$a_{1}$~ & ~$a_{2}$~ & ~$a_{z}$~ & ~$\Delta_{z}$~ & ~$E_{\mathrm{coh}}$~ & ~$d$~ \\ \hline
Bulk & Exp. & 4.37 & 3.31 & 10.47 & 2.16 & - & 3.07 \\
 & DZDP & 4.40 & 3.34 & 10.67 & 2.17 & 25.66 & 3.16 \\
 & ~~DZDP+vdW~~ & 4.36 & 3.34 & 10.38 & 2.17 & 26.49 & 3.02 \\
 & PAW & 4.54 & 3.31 & 11.17 & 2.12 & 21.43 & 3.46 \\ \hline
~Monolayer~ & DZDP & 4.44 & 3.32 & - & 2.15 & 25.27 & - \\
& DZDP+vdW & 4.43 & 3.32 & - & 2.15 & 25.78 & - \\
& PAW & 4.62 & 3.30 & - & 2.10 & 21.40 & - \\ \hline
\end{tabular}
}
\caption{The equilibrium lattice constants, $a_{1}$, $a_{2}$, and $a_{z}$ (in units of \AA), buckling $\Delta_{z}$, cohesive energy $E_{\mathrm{coh}}$ (in units of $\mathrm{eV}$), and interlayer distance $d$ (in units of \AA) for bulk and monolayer black phosphorus. Experimental data is reported in Ref.~[\onlinecite{ExpBBP}]}
\label{tab1}
\end{table}
%%%%%%%%%%%%%%%%%%%%%%%%%%%%%%%%%%%%%%%%%%%%%%%%%%%%%%%%%%%%%%%%%%%%%%%%%%%%%%%%%%%%%%%%%%%%%%%%%%%%%%%%%
%%%%%%%%%%%%%%%%%%%%%%%%%%%%%%%%%%%%%%%%%%%%%%%%%%%%%%%%%%%%%%%%%%%%%%%%%%%%%%%%%%%%%%%%%%%%%%%%%%%%%%%%%
%%%%%%%%%%%%%%%%%%%%%%%%%%%%%%%%%%%%%%%%%%%%%%%%%%%%%%%%%%%%%%%%%%%%%%%%%%%%%%%%%%%%%%%%%%%%%%%%%%%%%%%%%
\section{Numerical Results and Discussion}
In this section, we present our main numerical results based on first-principles simulations. Our aim is to explore the impact of strain on vibrational, mechanical and electronic properties of phosphorene. All the first-principles calculations are performed at room temperature.

\subsection{Elastic and vibrational properties of strained phosphorene}
\begin{figure}[t]
\centering
\includegraphics[width=\linewidth]{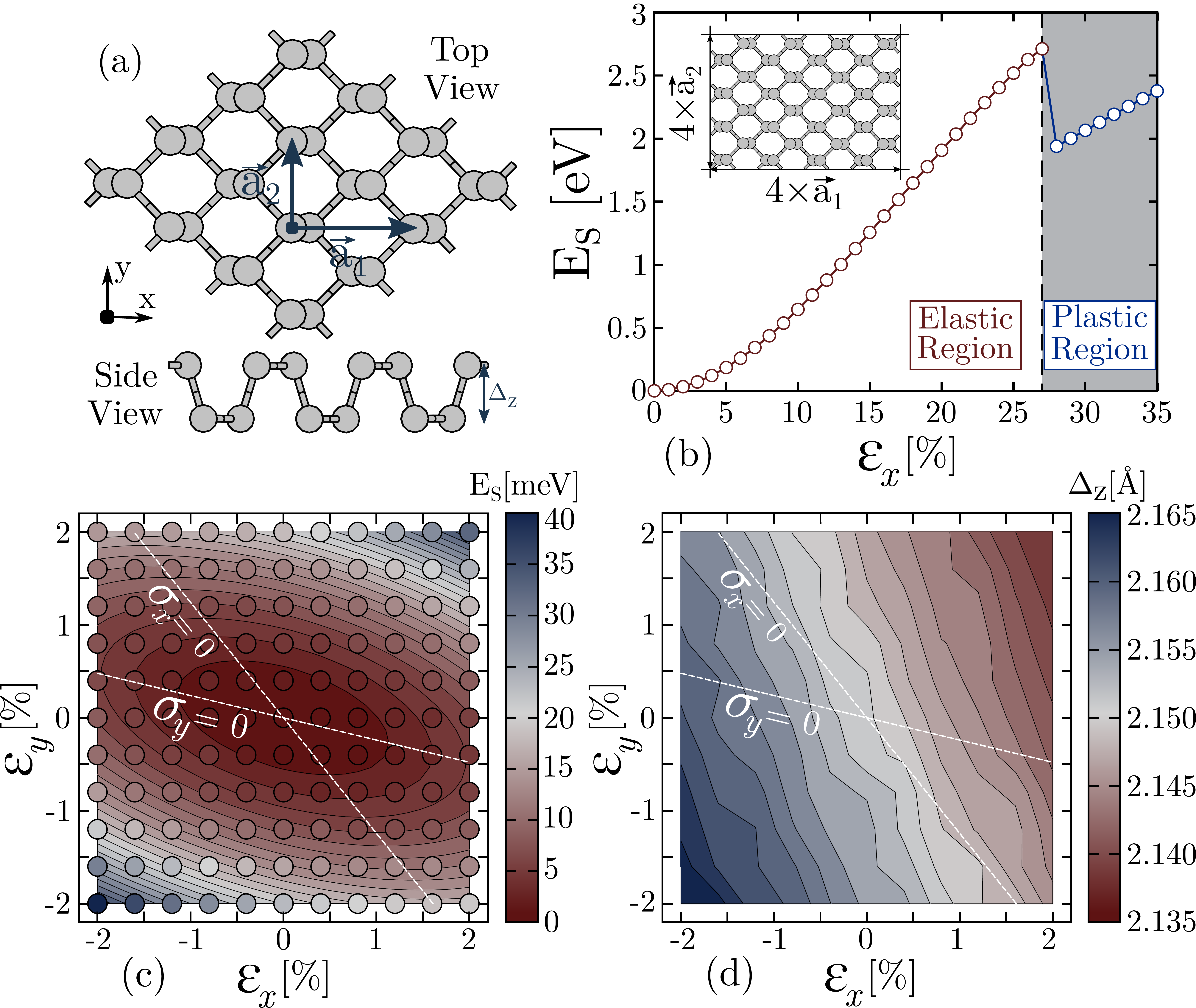}
\caption{\label{f:Fig1} (a) Schematic representation of the atomic structure of mono-layer phosphorene from the top and side views. (b) The per unit cell strain energy as a function of strain in uniform deformation regime. The shaded region indicates the plastic range. The inset shows the $4\times4$ rectangular supercell used in the calculations. (c) The surface plot of ($\varepsilon_{x}$, $\varepsilon_{y}$) and the corresponding per unit cell strain energies. The points denote actual data and the background is the fitted formula. (d) The mesh plot of $\Delta_{z}$ at the same data points as in (c).}
\end{figure}

In-plane lattice constants are either stretched or compressed by $\varepsilon_{x}$ and $\varepsilon_{y}$ in a $\mathrm{4\times4}$ supercell (see the inset of Fig.~\ref{f:Fig1}(b)) in order to obtain the strained structure. The consequent structure is then relaxed with keeping the deformed lattice vectors unchanged.

We obtain the variation in strain energy, $E_{S}$, by subtracting the total energy of the deformed structure from the equilibrium total energy, as the strain varies from 0\% to 35\% in the uniform expansion regime ($\varepsilon_{x}=\varepsilon_{y}\geq 0$). From Fig.~\ref{f:Fig1}(b), the harmonic region can be assumed within the strain range of 0-0.02 and afterwards the anharmonic region occurs and is basically followed by a plastic region (see the shaded area in Fig.~\ref{f:Fig1}(b)) where irreversible changes occur in the structure of the system. The corresponding yielding strain is found to be 27\%, which is similar to that reported for graphene and molybdenum disulphide (MoS$_2$), revealing the promise of phosphorene for stretchable electronic devices~\cite{YSg, YSM}.

We further calculate Poisson's ratio $\nu$, the ratio of the transverse strain to the axial strain, along with the in-plane stiffness parameters $C$, to assess the mechanical response of phosphorene. Figure~\ref{f:Fig1}(c) shows the mesh plot of strain ($\varepsilon_{x}$, $\varepsilon_{y}$) and the corresponding strain energies. The strain-energy relation is then obtained as $E_{S}=a\varepsilon_{x}^{2}+b\varepsilon_{y}^{2}+c\varepsilon_{x}\varepsilon_{y}$, where $a$, $b$, and $c$ are fitted parameters obtained as $14.88$, $50.06$, and $23.94$ eV, respectively. We then calculate stress along the $x~(y)$-direction, denoted by $\sigma_{x(y)}$ through $\sigma_{x(y)}=V^{-1}_{0}\partial E_{S}/\partial \varepsilon_{x(y)}$, where $V_{0}$ is the equilibrium volume. The dashed lines denoted by $\sigma_{x}\mathrm{=0}$ and $\sigma_{y}\mathrm{=0}$ shown in Fig.~\ref{f:Fig1}(c), correspond to uniaxial deformations along the $x$- and $y$-directions, respectively. The associated Poisson's ratios are evaluated as $\nu_{y}\mathrm{=}c/2a\mathrm{=}0.81$ and $\nu_{x}\mathrm{=}c/2b\mathrm{=}0.24$, in consistency with those obtained via VASP package in Ref.~[\onlinecite{Wei1}] ($\nu_{y}\mathrm{=}0.7$ and $\nu_{x}\mathrm{=}0.2$). In comparison with the isotropic Poisson's ratios reported for graphene, boron nitride (BN), and MoS$_2$ (i.e., 0.16, 0.21, and 0.25, respectively), phosphorene has larger Poisson's ratios along both the armchair and zigzag directions~\cite{APL, Mos2_strain}.

With $A_{0}$ as the equilibrium area of the system, the in-plane stiffness along the $x~(y)$-direction is defined as $C_{x(y)}=A^{-1}_{0}\partial^{2}E_{S}/\partial^{2} \varepsilon_{x(y)}$. The corresponding Poisson's ratio results in $C_{x}\mathrm{=}A^{-1}_{0}(2a-\frac{c^{2}}{2b})\mathrm{=}26.16\mathrm{J/m^{2}}$ and $C_{y}\mathrm{=}A^{-1}_{0}(2b-\frac{c^{2}}{2a})\mathrm{=}88.02\mathrm{J/m^{2}}$. These values are smaller than those values reported for graphene, BN, and MoS$_2$ (i.e., 335, 267, and 123 $\mathrm{J/m^{2}}$, respectively), implying that phosphorene is more flexible along both armchair and zigzag directions~\cite{APL, Mos2_strain}. It should be mentioned that the calculated parameters are in excellent agreement with those reported in Ref.~[\onlinecite{Wei2}] ($24.42$ and $92.13\mathrm{J/m^{2}}$, obtained by converting the given data using a thickness of 5.55\AA).

It is worth mentioning that BBP exhibits a negative Poisson's ratio along its armchair direction in response to perpendicular uniaxial strains~\cite{nPBBP}. In order to probe the existence of a similar behavior in phosphorene, Fig.~\ref{f:Fig1}(d) depicts the monolayer's thickness for the same set of strain components as in Fig.~\ref{f:Fig1}(c). Under uniaxial strain along the $y$-axis, the thickness is reduced (increased) as the sheet is compressed (stretched). This manifests the existence of a negative out-of-plane Poisson's ratio in response to $y$-oriented uniaxial deformations. The calculated out-of-plane Poisson's ratios are 0.21 and -0.09, for uniaxial strains along the armchair and zigzag directions, respectively. Employing SIESTA with a double-$\zeta$ basis set, a previous study~\cite{nPMP} has reported the out-of-plane Poisson's ratios to be $0.046$ and $-0.043$, respectively. Despite the inconsistency, the presence of a negative Poisson's ratio is revealed in both studies. Such discrepancy can be attributed to diverse methodologies and fitting procedures adopted for obtaining the Poisson's ratios. Another point which might have negatively impacted the accuracy of the calculated Poisson's ratios in Ref.~[\onlinecite{nPMP}] is that the obtained equilibrium lattice constants are 10\% larger than those previously reported in the literature~\cite{ACSNano, Wei1, Wei2}.

\begin{figure}
\centering
\includegraphics[width=\linewidth]{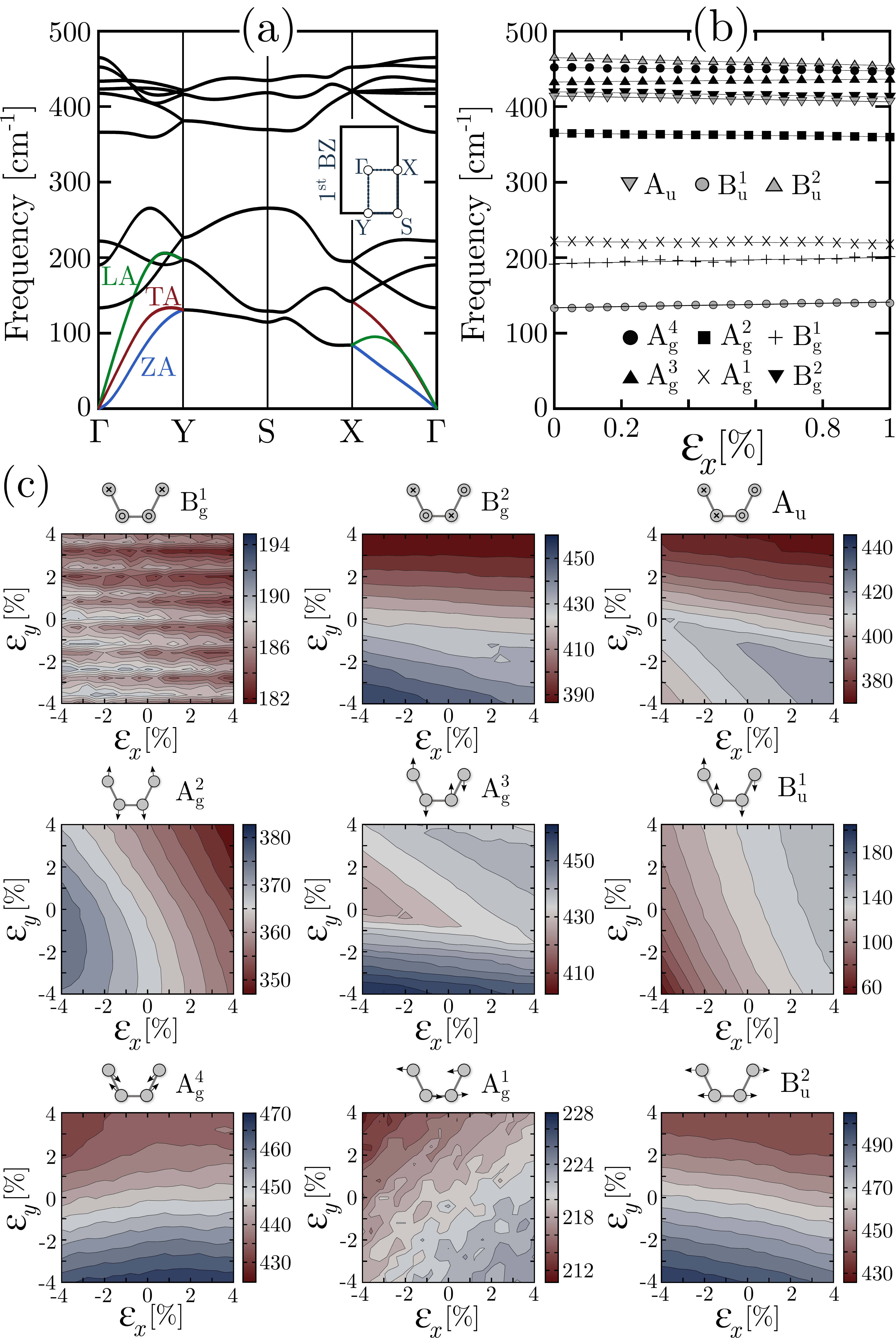}
\caption{\label{f:Fig2}  (a) The phonon dispersion curve of the undeformed phosphorene. ZA marks the out-of-plane acoustic branch and LA (TA) denotes in-plane longitudinal (transverse) acoustic vibrations. (b) Frequencies of the Raman/infrared active modes at the $\Gamma$ point of phosphorene under strain. (c) Contour plots of Raman/infrared frequencies with strain. The eigenvector of the corresponding vibrational mode is depicted at the top of each panel.}
\end{figure}

Phonon dispersions for an undeformed monolayer phosphorene are depicted in Fig.~\ref{f:Fig2}(a). To obtain the sound velocities, we calculate the slopes of in-plane acoustic branches in the vicinity of the $\Gamma$ point. The sound velocities in the $\Gamma-Y$ direction are derived as $7.59$ km/s and $4.48$ km/s for longitudinal and transverse atomic motions, respectively. Along the $\Gamma-X$ axis, on the other hand, the sound velocities are obtained as $5.69$ km/s and $5.27$ km/s for longitudinal and transverse vibrations, respectively. A previous study reports the respective maximum sound velocities along the $\Gamma-X$ and $\Gamma-Y$ paths~\cite{BlueP} as $3.8$km/s and $7.8$km/s, which is in good agreement with our results only along the $\Gamma-Y$ direction. We attribute the difference in the $x$-directed velocity to the instabilities observed in the out-of-plane acoustic phonon branch presented in Ref.~[\onlinecite{BlueP}], implying that the symmetry restrictions might be neglected during the phononic calculations. At 1\% uniform expansion, the respective sound velocities for LA and TA branches are equal to $7.6$ km/s and $3.31$ km/s along the $y$-direction and $3.77$ km/s and $6.04$ km/s along the $x$-axis.

Raman/infrared spectroscopy, as a versatile tool for structural characterization, has been widely used to study the electronic and vibrational properties in materials. The Raman spectrum is directly linked to the lattice dynamics of materials including phonon dispersion curves, phonon density of states, and infrared/Raman active modes. In accordance with the $C_{2h}$ point group symmetry of phosphorene, the modes $A_{g}^{1}$, $A_{g}^{2}$, $A_{g}^{3}$, $A_{g}^{4}$, $B_{g}^{1}$, and $B_{g}^{2}$ are characterized to be Raman active while the other three modes, namely, $A_{u}^{}$, $B_{u}^{1}$, and $B_{u}^{2}$, are infrared active. Figure~\ref{f:Fig2}(b) shows the frequency shifts of these optical phonon modes as a function of symmetric biaxial strain calculated with $\Gamma$-point only simulations. The Gr\"uneisen parameter, the variational frequencies of the individual atoms in phosphorene lattice varied with volume, for a vibrational mode $X$ ($\gamma_{X}$), is then calculated as $\gamma_{X} = - (2\omega^{0}_{X})^{-1} \partial \omega_{X}/\partial\varepsilon_{x}$, where $\omega^{0}_{X}$ is the frequency of mode $X$ in the absence of strain. The extracted average slope and the Gr\"uneisen parameter for all Raman and infrared active modes are presented in Tables \ref{tab2} and \ref{tab3}. As different modes exhibit qualitatively different behaviors in response to the applied strain, both negative and positive values for the slope and the Gr\"uneisen parameters are detected.

Figure~\ref{f:Fig2}(c) shows the contour plots of Raman/infrared frequencies with strain. The schematic representation of the atomic motions in each optical mode is also shown on top of each panel. As different modes demonstrate distinct trends under similar strain conditions, the frequency shifts of Raman/infrared active modes may serve as fingerprints of certain strain conditions, rendering them viable tools for mapping strain information from spectroscopy measurements.

\begin{table}[h]
\centering
\begin{tabular}{c|c|c|c|c|c}
\hline
Raman Active& \multicolumn{3}{c|}{$\omega_{X}$(cm$^{-1}$)} & \multicolumn{2}{c}{~~Gr\"uneisen Parameters~~} \\ \cline{2-6}
Modes ($X$) &~DZDP~ &~PAW~&~Exp.~& \multicolumn{1}{c|}{~$\partial \omega_{X}/\partial \varepsilon_{x}$~} & \multicolumn{1}{c}{~$\gamma_{X}$~}\\ \hline
$A_{g}^{1}$ & 221.52 & 221.65 & --- & -149.6 & 0.338 \\
$A_{g}^{2}$ & 365.09 & 341.88 & 363 & -465.7 & 0.638 \\
$A_{g}^{3}$ & 432.78 & 424.01 & --- & -688.5 & 0.821 \\
$A_{g}^{4}$ & 452.15 & 448.19 & 471.3 & -460.1 & 0.509 \\
$B_{g}^{1}$ & 181.43 & 192.67 & --- & 810.1 & -2.233 \\
$B_{g}^{2}$ & 413.65 & 424.52 & 440 & 373.9 & -0.432 \\ \hline
\end{tabular}
\caption{Phonon frequencies of the relevant Raman mode symmetry representations of phosphorene along with the corresponding Gr\"uneisen parameters. Our numerical results are compared with those results measured in experiment~\cite{Plasma}. }
\label{tab2}
\end{table}

\begin{table}[h]
\centering
\begin{tabular}{c|c|c|c|c}
\hline
\multicolumn{1}{c|}{~IR Active~} & \multicolumn{2}{c|}{$\omega_{X}$(cm$^{-1}$)} & \multicolumn{2}{c}{~Gr\"uneisen Parameters~} \\ \cline{2-5}
~Modes($X$) &~DZDP~&~PAW~& \multicolumn{1}{c|}{~$\partial \omega_{X}/\partial \varepsilon_{x}$~} & \multicolumn{1}{c}{~$\gamma_{X}$~}\\ \hline
$A_{u}$~($\vec{E}\mathrm{\parallel}\vec{a_{2}}$) & 419.30 & 416.53 & -761.4 & 0.920 \\
$B_{u}^{1}~$($\vec{E}\mathrm{\parallel}\vec{a_{z}}$) & 129.02 & 138.01 & 739.8 & -2.867 \\
$B_{u}^{2}~$($\vec{E}\mathrm{\parallel}\vec{a_{1}}$) & 465.32 & 457.49 & -1062.4 & 1.142 \\
\hline
\end{tabular}
\caption{Phonon frequencies of the relevant infrared mode symmetry representations of phosphorene along with the corresponding Gr\"uneisen Parameters. $\vec{E}$ is the polarization of the incident light.}
\label{tab3}
\end{table}

%%%%%%%%%%%%%%%%%%%%%%%%%%%%%%%%%%%%%%%%%%%%%%%%%%%%%%%%%%%%%%%%%%%%%%%%%%%%%%%%%%%%%%%%%%%%%%%%%%%%%%%%%
%%%%%%%%%%%%%%%%%%%%%%%%%%%%%%%%%%%%%%%%%%%%%%%%%%%%%%%%%%%%%%%%%%%%%%%%%%%%%%%%%%%%%%%%%%%%%%%%%%%%%%%%%
\subsection{Electronic Properties of strained phosphorene}
\begin{figure}
\centering
\includegraphics[width=\linewidth]{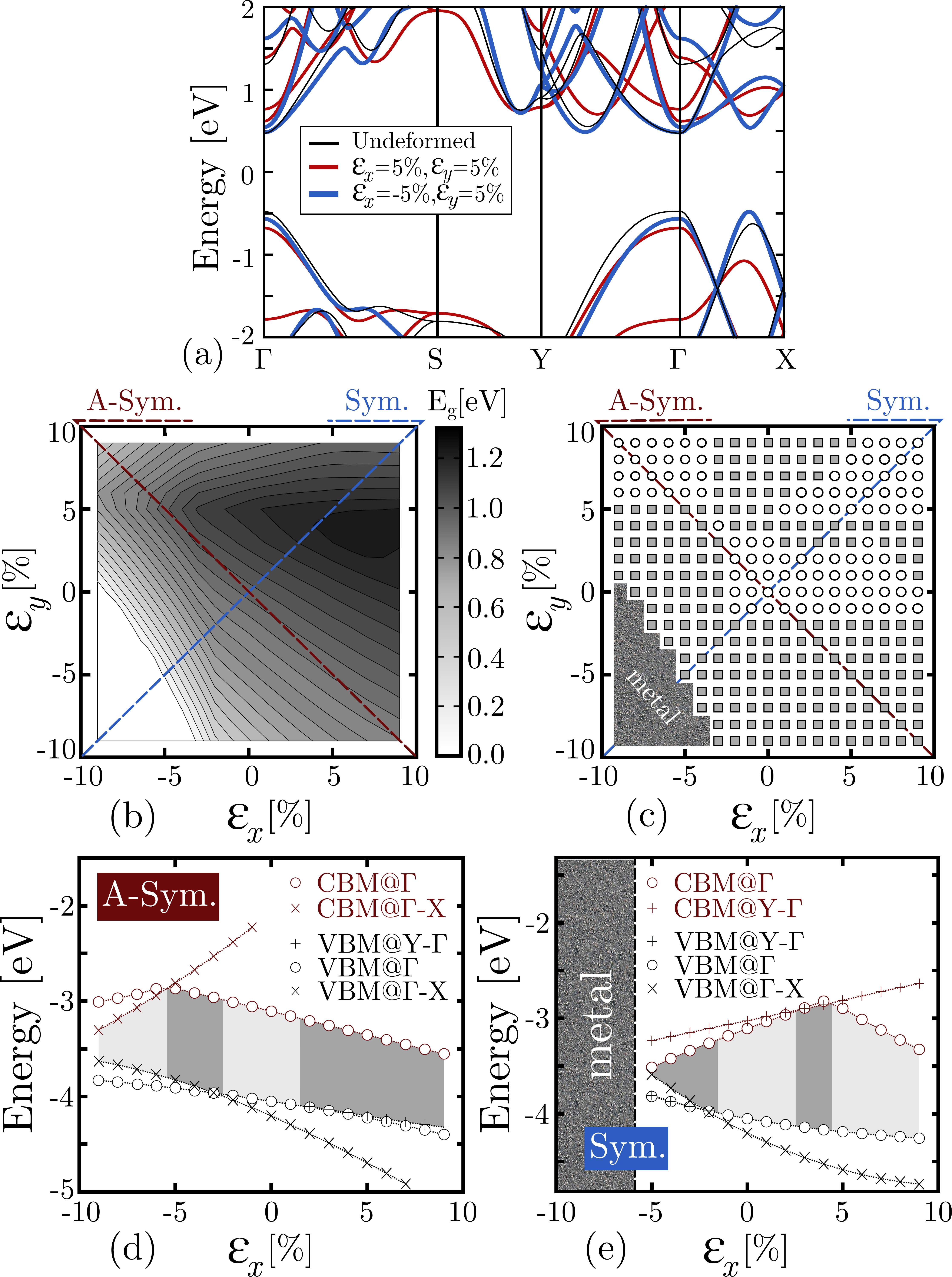}
\caption{\label{f:Fig3} (a) Modification of the electronic band structure under various strain configurations. (b) The surface plot of ($\varepsilon_x$,$\varepsilon_y$) and the corresponding band gaps. The two dashed diagonal lines denote the symmetric (Sym.) and anti-symmetric (A-Sym.) strain distributions. (c) Nature of band gaps for the same set of data presented in (b). Squares (circles) represent indirect (direct) band gaps. (d) and (e) show the VBM and CBM in symmetric and anti-symmetric strain distributions, respectively. Light (dark) gray regions correspond to direct (indirect) band gaps. Note that in (a) the Fermi energy is set to zero. In (d) and (e), the energies are referenced to vacuum level to further illustrate the modification of band offsets in strained structure.}
\end{figure}

In order to assess how two aspects of mechanical and electronic properties can be beneficially merged in the context of tunable electronic features, the electronic properties of monolayer phosphorene under various strain distributions are studied in this section. Figure~\ref{f:Fig3}(a) compares the electronic band structures of deformed phosphorene with its undeformed counterpart, along particular straight lines in $k$-space, according to which the substantial influences of strain on both band spacing and curvature are evident. Figures~\ref{f:Fig3}(b) and (c) illustrate the strain dependence of the size and nature of the band gap, respectively.

For the undeformed phosphorene, the band gap is calculated to be $0.95$ eV (0.91 eV)- as obtained using the SIESTA (VASP) package- in excellent agreement with previous studies~\cite{ACSNano, Castro}. Inspecting the nature of the band gap, our calculations based on both packages provide identical trends for the first conduction band in the vicinity of $\Gamma$-point, which is the exact position of the conduction band minimum (CBM). For the first valance band, however, the actual placement of valance band maximum (VBM) slightly differs from SIESTA to VASP. While the SIESTA band structure predicts the VBM to be located precisely at $\Gamma$-point, the VASP package suggests an indirect band gap with its actual valance maximum occurring along the $\Gamma-Y$ high-symmetry line, 0.0285$\times 2\pi/a_{2}$ away from the $\Gamma$-point. The discrepancy can be attributed to the different calculation methods employed, i.e. the pseudopotential scheme combined with atomic orbitals in SIESTA versus projector augmented wave formalism with plane waves in VASP. Based on our calculations, a very marginal change of the overlap between atomic orbitals would transform the nature of the band gap from direct to indirect and vice versa. In fact, a recent symmetry analysis on undeformed phosphorene has provided a criterion based on which the direct/indirect nature of the band gap can be determined \cite{P-Li}. The authors, however, mentioning the marginal discrepancy between the two cases, and further by referring to shortcomings ascribed to DFT-based calculations, did not provide a determined conclusion regarding the exact position of VBM. Thus, whether phosphorene is truly a direct or nearly direct semiconductor (as dubbed in Ref.~\cite{Castro}), we believe it should be left to experimental studies.

For deformed structures, the maximum attainable direct (indirect) band gap is evaluated to be 1.34 eV (1.37 eV) which occurs at $\varepsilon_{x}=6\%,\varepsilon_{y}=3\%$ ($\varepsilon_{x}=6\%,\varepsilon_{y}=4\%$). For the anti-symmetric case ($\varepsilon_{x}=-\varepsilon_{y}$) in the strain range under study ($-9\% \leqslant\varepsilon_{x},\varepsilon_{y}\leqslant 9\%$), no semiconductor-metal transition can be triggered and a direct-indirect-direct-indirect transition is observed in the band gap. For symmetric deformations ($\varepsilon_{x}=\varepsilon_{y}$), the band gap experiences an indirect-direct-indirect-direct transition with a semiconductor-metal transition through the application of compressive strains larger than 6\%. Figures~\ref{f:Fig3}(d) and (e) show the details of the variations in the location of the band gap for anti-symmetric and symmetric strain distributions, respectively. Accordingly, in the symmetric case, both CBM and VBM undergo transitions between $\Gamma$ and $\Gamma-X$, giving rise to four distinct strain zones with boundaries at -2\% , 2\%, and 4\%. Inspecting the anti-symmetric case, while CBM experiences a transition similar to symmetric deformations, VBM moves between $\Gamma$, $\Gamma-X$, and $Y-\Gamma$, resulting in four strain zones with boundaries located at -6\%, -3\%, and 1\%.

\begin{figure}
\centering
\includegraphics[width=\linewidth]{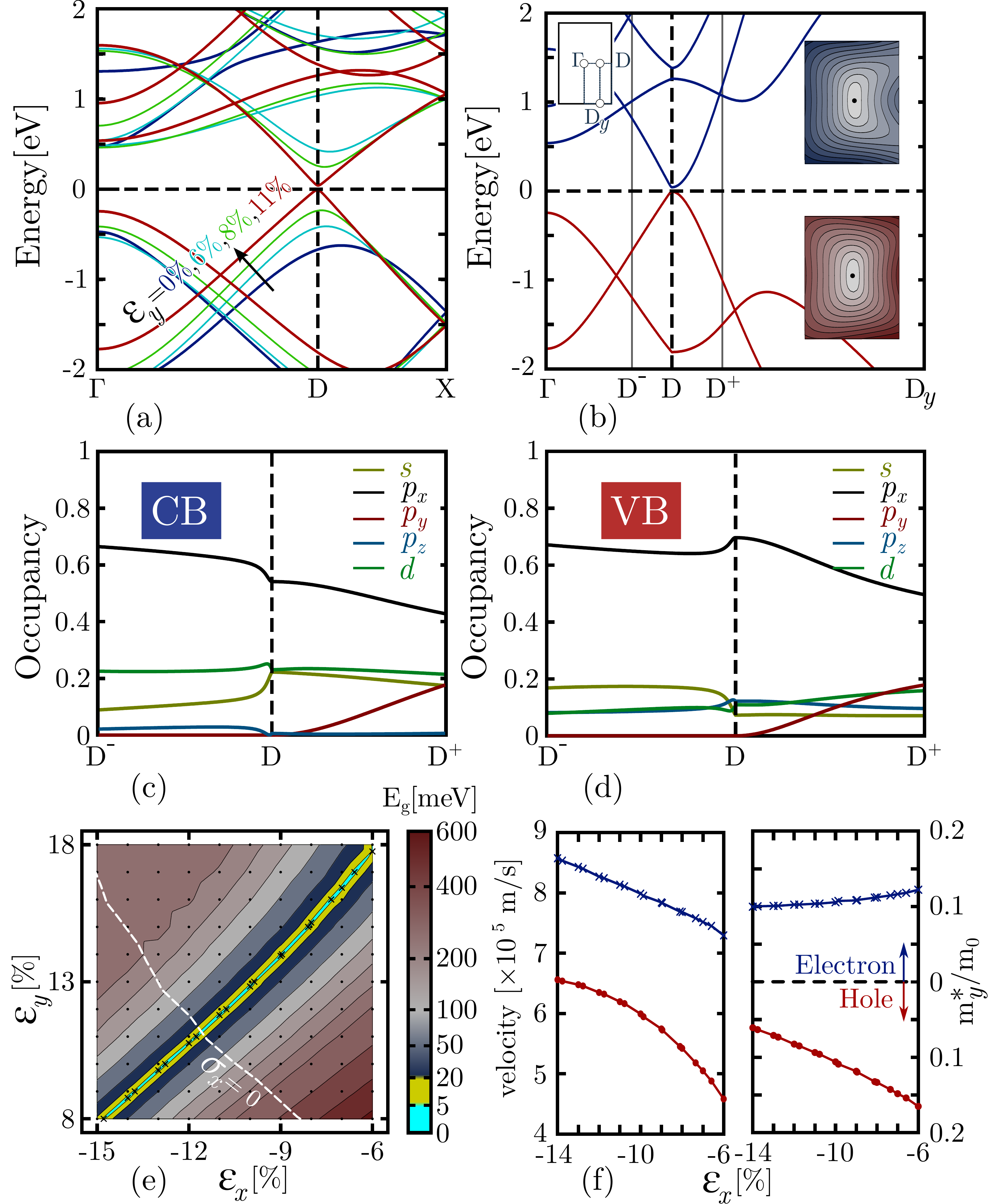}
\caption{\label{f:Fig4} (a) The band structures of phosphorene under various A-Sym. strain values and the emergence of a Dirac-shaped dispersion. Blue, cyan, green, and red denote $\varepsilon_{y}\mathrm{=-}\varepsilon_{x}\mathrm{=0\%,~6\%,~8\%,~and~11\%}$, respectively. (b) Zoom of the band structure at $\varepsilon_{y}\mathrm{=-}\varepsilon_{x}\mathrm{=11\%}$ along the selected paths of the first Brillouin zone. Inset shows the equi-energy contours for the conduction (top) and the valence (down) bands centered at $D$. Both rectangles span a length of $0.02\times\pi/a_{x}$ ($0.02\times\pi/a_{y}$) along the $k_{x}$ ($k_{y}$) direction of the first Brillouin zone. (c) and (d) denote the orbital composition of the crystal wave-functions close to the Dirac point for the topmost valence and the lowest conduction bands, respectively. (e) The mesh plot of the band gap at data points ($\varepsilon_x$,$\varepsilon_y$). Bright cyan (yellow) region denotes Dirac-shaped dispersions with band gap smaller than 5 (20) meV. The cross marks denote selected strain distributions for which the existence of a Dirac-like dispersion is further validated by VASP. (f) Fermi velocity and $y-$directed effective mass for Dirac-shaped dispersions as functions of $\varepsilon_{x}$. Crosses and filled circles denote the results for conduction and valence energy bands, respectively.}
\end{figure}

Figure \ref{f:Fig4}(a) shows the variation of the band structure under anti-symmetric strain. As seen, the band gap has almost vanished at $\varepsilon_{y}\mathrm{=}-\varepsilon_{x}\mathrm{=11\%}$ and a linear dispersion emerges at the $D$ point. Figure~\ref{f:Fig4}(b) shows the energy dispersion along both the $\Gamma-D$ path and the direction perpendicular to it in the vicinity of $D$ point. For both conduction and valence bands, despite the linearity along $\Gamma-D$, the bands are parabolic along the $D-D_{y}$ path. In addition, the associated slope (curvatures) of the conduction and valence bands along the $\Gamma-D$ ($D-D_{y}$) direction are remarkably different. This can be further approved by referring to the energy contours for both conduction and valence bands (see the insets of Fig.~\ref{f:Fig4}(b)). Although the D point by itself is no longer a high symmetry point, it still lies along lines of fairly high symmetry in the Brilloun zone.
Decompositions of the valence and conduction band crystal wave-functions around $D$ over the constituent atomic orbitals $s$, $p_{x}$, $p_{y}$, $p_{z}$, and $d$, are depicted in Figs.~\ref{f:Fig4}(c) and (d) for the selected strain of $\varepsilon_{y}=-\varepsilon_{x}\mathrm{=11\%}$. Symmetry of the system mandates equal contributions to the crystal wave-function from all four atoms in the unit cell of phosphorene. For each atom, despite the prevalence of $p_{x}$, the contributions from other orbitals, especially those of $s$ and $d$, should also be taken into account to properly describe the linear energy branches.

To unravel the formation possibility of similar Dirac-shaped dispersions via strain, we performed a thorough inspection of band gaps, employing a dense network of data points ($\varepsilon_{x}$,$\varepsilon_{y}$) shown in Fig.~\ref{f:Fig4}(e). In this figure, the bright cyan (yellow) area denotes Dirac-shaped dispersions with band gaps smaller than 5 (20) meV. Our calculations show the attainability of Dirac-like spectrum via invoking uniaxial deformations parallel to the zigzag axis. To illustrate this, the dashed line pertaining to $\sigma_{x}=0$ is superimposed on the mesh. The cross marks on Fig.~\ref{f:Fig4}(e) denote selected strained lattice vectors for which the existence of a Dirac-like dispersion is further authenticated by VASP. The maximum discrepancy between band gaps obtained from VASP and SIESTA in the selected geometries is 12.62 meV. We therefore conclude that, as far as DFT based simulations are concerned, our prediction regarding the existence of a Dirac-like dispersion is valid. It should be mentioned that the asymmetric strain of $\varepsilon_{y}=-\varepsilon_{x}=11\%$, shown in Fig.~\ref{f:Fig4}(b), is provided as a sample exterior to the cyan region of Fig.~\ref{f:Fig4}(e), which still clearly manifests the Dirac-like feature. According to Fig.~\ref{f:Fig4}(e), although the band gap opens up beyond the cyan region, our calculations shows that the anisotropic Dirac-liked energy spectrum remains intact for band gaps of up to 55meV. For the emergence of Dirac-like spectrum deformations as large as 11$\%$ might be needed. There are now various practical schemes on how to incorporate strain into a 2D material. It has been reported that flexible substrates can be used to apply tensile axial strains of up to 30$\%$ to a graphene sheet \cite{ahn}. Moreover, as the corresponding tensile yielding strains are calculated to be 27$\%$ and 30$\%$ along the zigzag and armchair axes, respectively, one can conclude that for strain magnitudes of up to 11$\%$, phosphorene sheet will experience no detrimental plastic deformation and thus will preserve its structural integrity \cite{Wei1, Wei2}. Hence, the Dirac-like feature can definitely lend itself to experimental verifications and practical applications.

Figure~\ref{f:Fig4}(f) illustrates the linear velocity, $v_k$, and $y-$directed effective mass of Dirac-shaped dispersions as a function of $\varepsilon_{x}$. The calculated effective mass of carriers for both conduction and valence bands illustrates the parabolic nature of the energy spectrum along $D-D_{y}$ for all the considered strain magnitudes. The variation in the effective mass along $D-D_{y}$ is in the range of $0.10-0.12$ and $0.06-0.17$  (in units of $m_{0}$) for the conduction and valence bands, respectively. $v_{k}$ along $\Gamma-D$ spans a range of $7.3-8.6\times10^{5}$ and $4.6-6.6\times10^{5}$ for the conduction and valence bands, respectively. For the linear branches along $\Gamma-D$, it can be seen that the associated velocity of the conduction band is at least $\times 1.3$ larger than that of the valence band, irrespective of the strain value. Noting that the Fermi velocities calculated for graphene, silicene and germanene, are $6.3\times10^{5}$, $5.1\times10^{5}$, and $3.8\times10^{5}$ m/s, respectively, it can be concluded that the Dirac-shaped dispersion of phosphorene is absolutely competitive with that of previously studied materials~\cite{GSiGeDirac}.

It is worthwhile to mention that a similar Dirac like dispersion has also been reported for 6,6, 12-graphyne, which has a rectangular crystal lattice~\cite{graphynePRL, graphyneAPL}. Of the two anisotropic Dirac cones in the first Brillouin zone of this material, the first one shows linear dispersion with the Fermi velocities of $v_{kx}\mathrm{=}4.9\times10^{5}$ m/s and $v_{ky}\mathrm{=}5.8\times10^{5}$ m/s, while the second one is parabolic near the center of the cone. Moreover, the maximum attainable Fermi velocity of deformed 6,6, 12-graphyne is nearly $6.6\times10^{5}$ m/s when uniaxially strained about 7\% along the $x$-axis. In comparison, phosphorene is different as it has distinct velocities pertaining to each of the two linear branches crossing at the $D$ point, with both being more adjustable via in-plane strain engineering.

%%%%%%%%%%%%%%%%%%%%%%%%%%%%%%%%%%%%%%%%%%%%%%%%%%%%%%%%%%%%%%%%%%%%%%%%%%%%%%%%%%%%%%%%%%%%%%%%%%%%%%%%%
%%%%%%%%%%%%%%%%%%%%%%%%%%%%%%%%%%%%%%%%%%%%%%%%%%%%%%%%%%%%%%%%%%%%%%%%%%%%%%%%%%%%%%%%%%%%%%%%%%%%%%%%%
%%%%%%%%%%%%%%%%%%%%%%%%%%%%%%%%%%%%%%%%%%%%%%%%%%%%%%%%%%%%%%%%%%%%%%%%%%%%%%%%%%%%%%%%%%%%%%%%%%%%%%%%%
\section{Conclusion}
In conclusion, a highly anisotropic mechanical response of phosphorene is revealed through the calculation of in and out-of-plane elastic constants. In particular, a negative out-of-plane Poisson's ratio is observed for uniaxial deformations along the zigzag direction. Compared to graphene and two-dimensional molybdenum disulphide, phosphorene is shown to possess a smaller (larger) in-plane stiffness (Poisson's ratio) along both armchair and zigzag axes while offering comparable yielding strength.

The vibrational frequencies of phosphorene are calculated and the corresponding shifts are obtained in response to various biaxial strain distributions. With the ability of detecting Raman/infrared frequency shifts via high resolution Raman/infrared spectroscopies, our results are of paramount importance for the characterization and mapping of strain distributions in phosphorene samples.

By inspecting various strain distributions, it is shown that in-plane deformations strongly affect the size and nature of the band gap. In addition, strain is shown to significantly modulate the effective mass of both electrons and holes in phosphorene.

Furthermore, we found that for specific deformations, including the $y$-oriented uniaxial tension, a linear energy spectrum with linear velocities comparable to those of other 2D semi-metal materials can be attained. The Dirac-like dispersion of deformed phosphorene, however, is distinct from those previously reported for graphene, silicene, and germanene, as in phosphorene the anisotropic dispersion allows carriers to behave as either massless Dirac fermions or massive charges, depending on the transport direction along the armchair or zigzag axes, respectively. Such an anisotropy in the linear velocity may trigger a corresponding direction dependence in resistance, rendering phosphorene as a promising candidate for future nano-electronic device applications. It is highly desirable that we are able to manipulate the electronic structure of phosphorene via strain engineering as it increases a number of potential applications in nano-electromechanical as well as nano-optomechanical systems.

{\it Note added--} During the last stage of preparing this manuscript, Ref.~[\onlinecite{LVM}], where authors reported a substantial shift of Raman peaks via strain engineering in phosphorene, appeared in arXiv.

\appendix

\section{Variation of mass with strain}

\begin{figure}
\centering
\includegraphics[width=\linewidth]{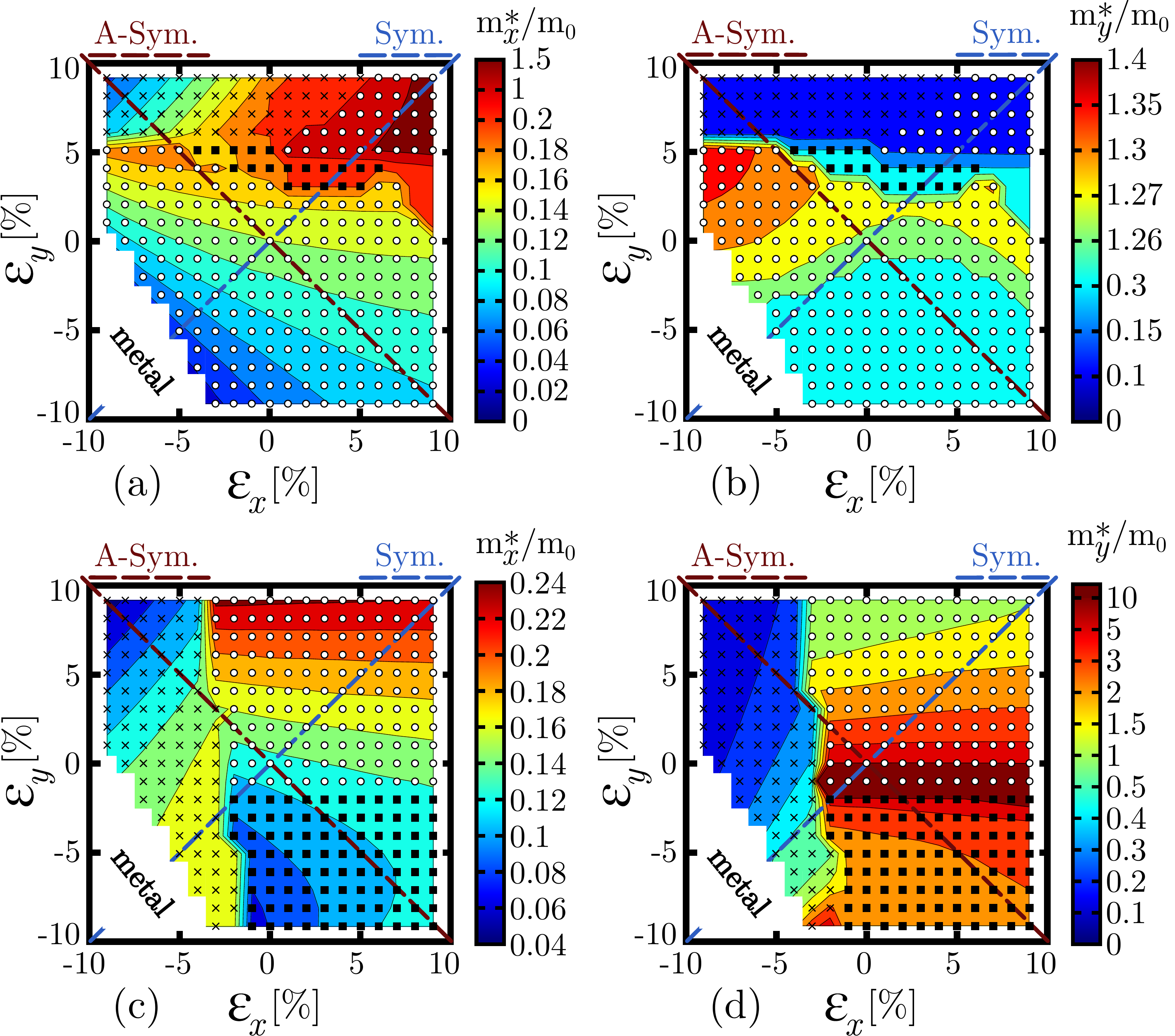}
\caption{\label{f:Fig5}(a) and (b) ((c) and (d)) depict the effective mass of electrons (holes) along the $x$- and $y$-directions, respectively. The corresponding locations of the VBM and CBM are denoted by circles (at $\Gamma$-point), crosses (along the $\Gamma-X$ high-symmetry line), and squares (along the $\Gamma-Y$ high-symmetry line).}
\end{figure}

Here, for the sake of completeness, we report our results for the variation of the effective masses of both electrons and holes in all the strain configurations. The mesh plot of the effective masses at data points $(\varepsilon_x,\varepsilon_y)$ are shown in Fig.~\ref{f:Fig5}. In addition, in this figure we denote the exact locations of the corresponding CBM (VBM) in which the effective mass of electron (hole) is calculated. As shown, the applied strain can widely tune the effective mass of carriers. The discontinuities in the values of effective masses are also found to be at strain values in which direct-indirect band gap transitions take place.

\bibliography{ref}

\end{document}